# Toward Development of a New Health Economic Evaluation Definition

**Alexei Botchkarev**

Principal, GS Research & Consulting
Adjunct Professor, Ryerson University
Toronto, Ontario, Canada.
Email: *alex.bot@gsrc.ca*

## Abstract

Economic evaluation (EE) is a dynamically advancing knowledge area of health economics. It has been conceived to provide evidence for allocating scarce resources to gain the best value for money. The problem of efficiency of investments becomes even more crucial with advances in modern medicine and public health which bring about both improved patient outcomes and higher costs. Despite the abundance of literature on the EE concepts, some key notions including the definition of the health economic evaluation remain open for discussion. Academic literature offers a large number and growing variety of EE definitions. It testifies to the fact that existing definitions do not meet economists' requirements. The aim of this study was to examine existing definitions and reveal their common features.

## 1   INTRODUCTION

Advances in modern medicine and public health bring about both improved patient outcomes and higher costs. Health care spending is growing in many countries. Projected US health annual spending growth for 2014-24 is to average 5.8 percent, raising the health share of US gross domestic product to 19.6 percent in 2024 (Keehan, et al, 2015).

Economic evaluations (EE) have been conceived to provide evidence for allocating scarce resources to gain the best value for money. Through its several decades of development, EE turned into an established multidisciplinary area of knowledge which lies on the intersection of economics, medicine, health care, evaluation research, outcomes research, comparative effectiveness research, health technology assessment, etc. (see Figure 1).

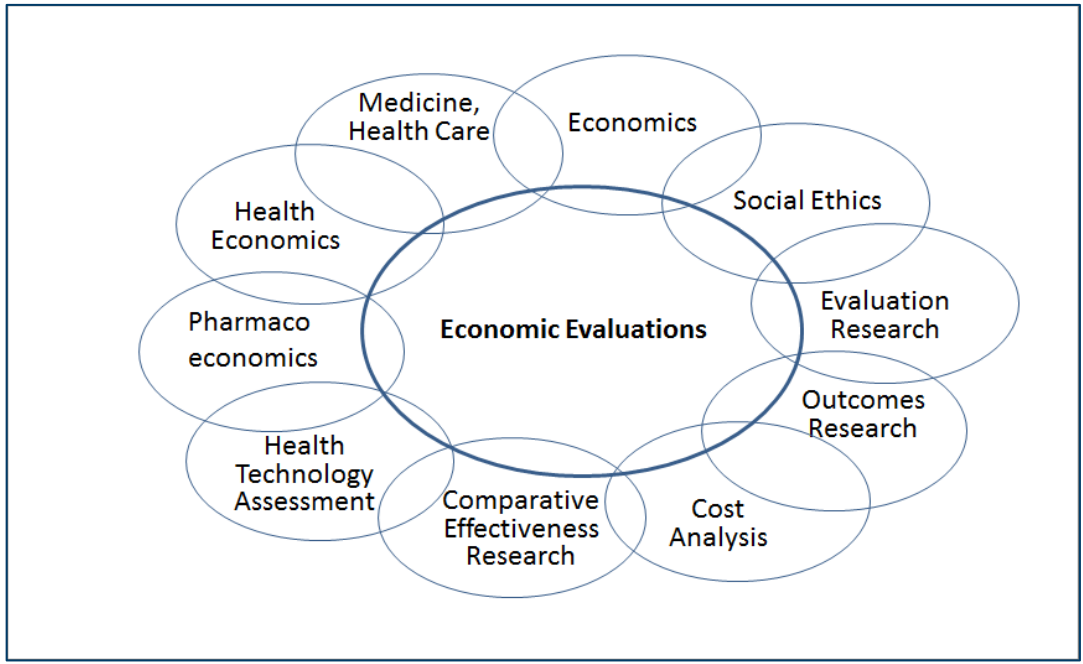

Figure 1. Economic evaluation lies on the intersection of subject areas of multiple sciences and fields of knowledge

EE is a dynamically advancing area of health economics. Increasingly large number of EEs is being conducted and reported. Figure 2 demonstrates the number of EEs published in academic journals over five-year periods from 1960 to 2020. The number of articles indexed in the Medline Complete database demonstrates the third order polynomial growth. See Note 1 below for the data collection method used to generate the graph.

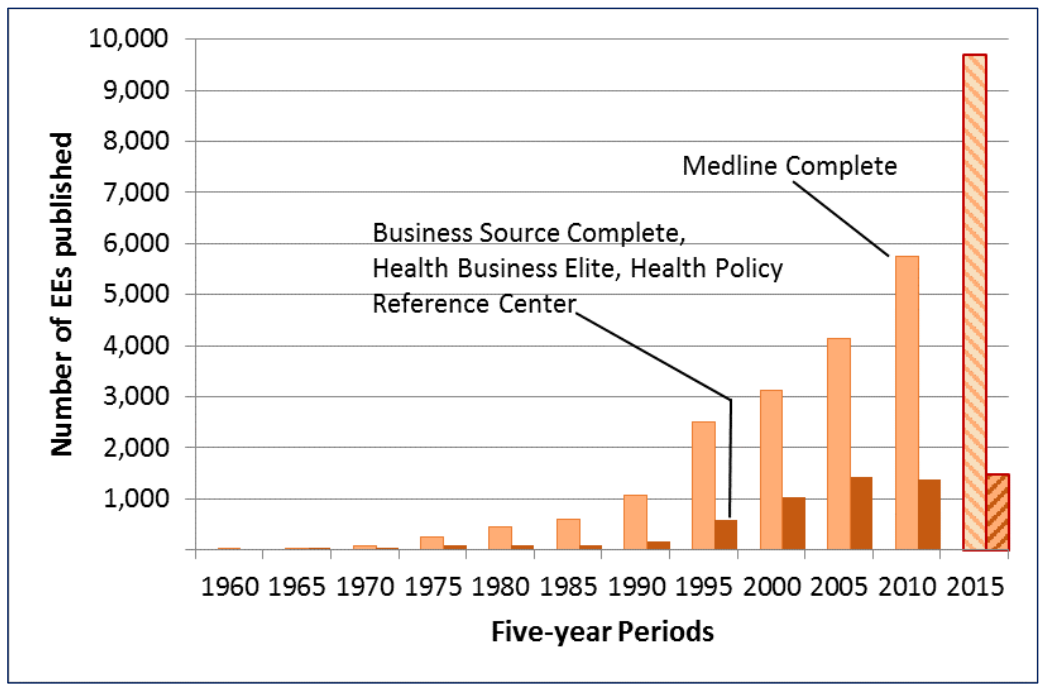

Figure 2. Historical dynamics of the EEs published in academic journals

Recent bibliographic analysis of the EE publications confirms global interest to and large volume of publications on the subject (Pitt, Goodman and Hanson, 2016). Pitt and colleagues identified 2,844 unique, complete EEs published during a period from 1 January 2012 to 3 May 2014 (28 months). These EEs were published in a total of 967 different journals (although 75% of these journals published no more than 1-2 articles). Based on the Pitt et al (2016) data, Figure 3 shows annual average numbers of publications by health area. Most intensive research is focused on five areas: cancer and other neoplasms, cardiovascular diseases, mental health, cognition, and developmental and behavioural disorders (including self-harm and substance disorders), musculoskeletal diseases (including back pain) and respiratory diseases – comprising over 60% of total EEs. In each of these health areas one to two hundred EEs published annually. EEs were also conducted in another 20 health areas. Average annual number of complete EEs was observed to be 1,219 (in 2012-14).

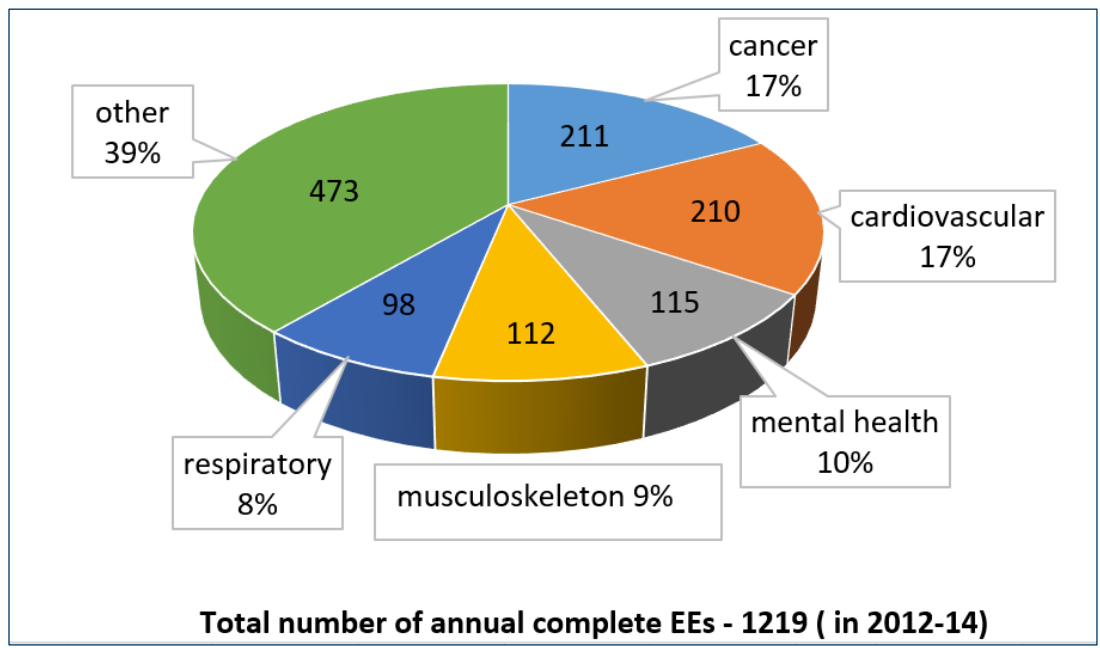

Figure 3. Annual number of EEs published in academic journals by health area

Even though the numbers of the EEs conducted around the world are impressive, most likely they are not all-inclusive. One of the reasons is a variability of the terminology used. In some cases EEs are referred to as economic appraisals (e.g. Bowling, 2014), technology appraisals (e.g. NICE, 2013) or economic assessments (e.g. Toumi, et al, 2015).

Despite the abundance of literature on the EE concepts, one of the key notions – EE definition – remains open for discussion. This is confirmed by multiple peer-reviewed papers offering a variety of EE definitions.

The purpose of this study is to explore existing EE definitions.

Literarature review was used as a methodology.

## 2   LITERATURE REVIEW

A widely accepted definition of economic evaluation is given by Drummond M.F. and his co-authors:

> "[Economic evaluation -] the comparative analysis of alternative courses of action in terms of both their costs and consequences." (Drummond, et al, 2015).

A number of EE definitions listed below clearly have been influenced by the Drummond's wording.

A plethora of healthcare economic evaluation definitions may be found in academic literature. Using Google Scholar search engine, we identified 60 non-identical definitions of health economic evaluation. The authors are attempting to reveal most important characteristics of the EE and combine them in a compact wording. Some definitions are intended to fully describe the notion, others focus only on certain aspects relevant to the authors' studies (e.g. highlighting EEs purpose, intention, goals or objectives). We analyzed and systematized all definitions based on the common features they demonstrate. A total of 8 groups were identified and are presented below. For better visualization, key words in each definition that prove association with a certain group have been italicized (all italic font in the quoted definitions has been done by the author of this paper, except for one case which has an explicit note).

Several authors emphasize that EE is based on the foundational principles of welfare economics. Specific economic concepts and terms are used in the definitions such as opportunity costs, resource consumption and production, monetary value, etc.

> "Economic evaluation is *a tool derived from the principles of welfare economics* to help decision makers make the best of limited resources… Economic evaluations use economic theory to develop a systematic framework by which to assess the relative costs and consequences of different interventions." (Kumaranayake, 2002).

> "*Firmly rooted in welfare economics*, economic evaluation is a practical tool for deciding whether a policy intervention will enhance social welfare or not, taking into account all (incremental) costs and benefits of that policy." (Brouwer, et al, 2007).

> "An economic evaluation is a comparison of outcomes among alternative ways of achieving common objectives. These analyses are conducted according to explicit, systematic and consistent criteria, and take into account both the positive and negative consequences of each alternative. Consequences may include economic, clinical, and humanistic outcomes… Economic outcomes represent the *consumption and production of resources and their monetary value* from the perspective of a decision maker." (Herman, Craig and Caspi, 2005).

> "Economic evaluation is a way of providing information about whether the benefits of adopting a course of action compensate for the *loss of benefits* which would have been generated had the best alternative use been made of the limited resources. It does this by undertaking a comparative analysis of alternative courses of action in terms of both their costs (resource use) and benefits." (Vale, et al, 2007).

> "The purpose of conducting economic evaluation in health care is an attempt to *determine the preferences of society* for a certain program/therapy in relation to the adopted health policy alternatives." (Suchecka and Jewczak, 2014)

"Economic evaluations assess the *level of input relative to the level of output* from a course of action. In health care, a course of action is a health intervention; the input or cost of an intervention could contain all *consumed resources* such as people, time, facilities and materials; the output (or consequence) could be any of effect of the intervention such as clinical end points, mortality, years of life or quality of life. Because costs are most time relative costs instead of absolute costs, at least two alternative health interventions are required in a full economic evaluation for the robustness of the comparison conclusion." (Zhang, 2009).

A clear trend of many definitions is an inclusion of an overarching EE goal - to provide information on which technologies will maximize value for money in healthcare.

"A health economic evaluation is a way of *establishing the "value for money"* of different health care technologies." (Kobelt study as cited in Tomson, 2005).

"Health economic evaluation is a method for evaluating the *value for money* of medical technologies by comparing alternative options in terms of their costs and consequences." (Hiligsmann, 2010).

"Economic evaluation is a rational approach for informing … decisions about what technologies or interventions should be prioritised over others *to maximize the value for money* in health care (and elsewhere)." (Søgaard, Kløjgaard and Olsen, 2010).

"Economic evaluation is *a search for value*. The objective: to ascertain the value of what is done in the name of public health in comparison to the value of what is not done, to help ensure that we make the best use of scarce resources." (Shiell, 2007).

"Economic evaluation is a tool to assist those responsible (governments, clinicians, administrators) to make rational decisions about new medical interventions, to try to understand what gives *good value for money* in the health service." (Jönsson, 2004).

"Economic evaluations compare the costs and consequences of two (or more) health care interventions. Economic evaluation is a way of thinking, backed up by a set of tools, which is designed *to improve the value for money* from investments in health care and welfare." (Fox-Rushby and Cairns, 2005).

In some definitions, the same idea is expressed using a concept of economic efficiency.

"Health economic evaluation aims at *providing information on the efficiency* of interventions." (Brousselle and Lessard, 2011).

"Economic evaluation of health care procedures and technologies is about *assessing their efficiency*, that is the produced health effects are weighed against the sacrifices or costs required attaining them. Efficiency is thus defined as a relationship between health effects and costs. Economic evaluation deals with establishing the *efficiency of the whole treatment process compared to another treatment process*." (Italics by the authors of the quoted study) (European Glaucoma Society, 2008).

"Economic evaluation of health care programmes aims to aid decision-makers with their difficult choices in allocating health care resources, setting priorities and moulding health policy. But it might be argued that this is only an intermediate objective. The real purpose of doing economic evaluation is *to improve efficiency*: the way inputs (money, labour, capital etc.) can be converted into outputs (saving life, health gain, improving quality of life, etc.)". (Miller, 2009).

"Economic evaluation is a form of comparative analysis; it allows interventions to be assessed in terms of their benefits and costs to provide information *to allocate resources efficiently*. … Economic evaluation facilitates comparisons between health care programmes, treatments, services and interventions in terms of both costs and consequences of those interventions." (Edlin, et al, 2015).

"Economic evaluation is a way in which we can attempt to allocate money to health care interventions in *the most efficient way* possible, gaining maximum health outcome with restricted investment." (Boyers, et al, 2015).

"Economic evaluation is a general term which refers broadly to a number of techniques designed to calculate the cost and effects of alternative resource allocations, and *to increase efficiency* in health service delivery". (Andrews, 2005).

A number of authors emphasize the intended purpose of the EE - to help policy and decision makers. This approach can be considered an element of systems thinking. Conducting EE and presenting a report is not an end-point. EE is a component of a larger system and outputs of the EE are intended for further analysis by policy and decision makers. Two important implications follow. First, results of the EE should contain economic evidence which fits (by content and format) the needs of the decision makers to have chances for implementation. Second, results of the EE (however crucial they are) constitute only one of the inputs (types of evidence) when decisions are made. Other inputs include quality, equity, public interest, policy preferences, etc.

"Simply put, economic evaluation is the understanding and *use of economic evidence in decision making*." (Rabarison, et al, 2015).

"Economic evaluation is a *crucial component of evidence-based policy-making*, helping identify the most efficient policies." (Pitman, 2012).

"Economic evaluation is *a technique to assist decision makers* in choosing between different courses of action or investment to maximize the health gain from any given budget." (Suh and Han, 2008).

"Health economic evaluation is a method used to analyse the costs and benefits of different health care interventions, and has often been quoted as the *most promising tool to assist decision-makers* in health care rationing." (Teerawattananon and Russell, 2008).

"Economic evaluation is *a tool that has been used as a decision-making* aid for optimal societal resource allocation. It provides a framework for explicitly measuring and comparing the value of alternative medical interventions in terms of their clinical, health-related quality-of-life, and economic outcomes." (Ackerman, Mafilios and Polly Jr, 2002).

"An economic evaluation is a method to compare outcomes and costs of interventions … with the *aim to improve resource allocation decisions by policy makers and insurers*." (van den Biggelaar, et al, 2011)

"Economic evaluation is *a tool used by policy and decision makers* to address the relationship between clinical effects and costs associated with diagnosis, treatment, adverse effects, supportive health care, and life gained or lost." (Ferrusi, et al, 2011).

"Economic evaluation is *a tool for assisting decision-making* given assumptions about how society wishes to maximize the benefits from limited health care spending. Economic evaluation involves identification of the benefits and costs of such spending, where: the benefits include improvements in the maintenance of, or prevention of deterioration in, health status (including improvements in length of life, reductions in illness and improvements in quality of life); and the costs are the resources that are used to generate the benefits." (Bhowmik, et al, 2014).

"Economic evaluation is a component of health economics, where the costs and consequences of alternative courses of action are compared to *provide evidence to help policy makers* and healthcare planners make these decisions". (Williams' book as cited in Kuper, Jofre-Bonet and Gilbert, 2006).

"Economic evaluation is a decision-making *tool for informing clinical practice and health policy…*". (Ford, et al, 2010).

"Economic evaluations provide decision makers with information on the tradeoffs in resource costs and public health benefits involved in choosing one intervention over another. …Economic evaluations, *along with research on safety, efficacy, and effectiveness* are crucial inputs for assessing health technologies and interventions for use in low-resource settings". (Levin, 2013).

"Health economic evaluation *is a dimension of health-technology assessment* that aims to inform decision-makers of the most efficient ways to use healthcare resources by weighing the benefits of a technology against its costs." (Rovithis, 2009).

"…economic evaluation is *a tool to support decision-making* that should be considered *alongside other evidence,* including issues of equity, generalizability, study quality and policy or public preference." (Byford and Barrett, 2010).

"At its core, economic evaluation is *a process designed to produce an estimate rather than a decision*, and it is likely that health care decisions will always be based on other factors in addition (perhaps) to economic efficiency." (Hoch and Dewa, 2005).

A number of authors point to an important characteristic of EE – its structured and systematic approach. Official agencies in many countries issued regulatory EE guidelines which are mandatory for use.

"Economic Evaluation. The *systematic appraisal* of costs and benefits of projects, normally undertaken to determine the relative economic efficiency of programs." (NICHSR, n.d.)

"Economic evaluation of health services is a branch of economics that deals with "*systematic evaluation* of the benefits and costs arising from the comparison of different health technologies." (Fragoulakis et al, 2015).

"Health economic evaluation comprises the *systematic appraisal* of the costs, the benefits and the relative economic efficiency of different medical interventions." (Lambert and Hurst, 1995).

"Economic evaluation is *a structured method* for examining the costs and consequences involved with alternative methods of treatments and/or programs, in order to inform which is the best alternative from a particular viewpoint." (Ford, 2012).

It is common for the authors to refer to EE as a framework (container or umbrella) concept encompassing a set of tools, methods or techniques for gathering and processing standardized and quantitative data.

"Economic evaluation is *a general umbrella term* that refers to a collection of analytic techniques all aimed at facilitating decision making regarding the overall merit of healthcare programs considering both costs and consequences." (Torrance, 1997).

"Health economic evaluation is a '*container concept*', embodying a broad range of different approaches. First, there are partial evaluations, which provide information on the cost implications of illnesses and interventions — albeit not from an efficiency perspective. For instance, cost-of-illness studies … Full economic evaluations compare the costs and effects of two (or more) competing options, often a new intervention versus an already implemented alternative." (Luyten, Naci, and Knapp, 2016).

"Economic evaluation is a systematic and transparent *framework* for assessing benefits; it is used to help make decisions and does not make decisions directly." (Lenoir-Wijnkoop, et al, 2013).

"Economic evaluation. A comparative analysis of at least two health interventions used to assess both the costs and consequences of the different health interventions, *providing a decision framework.*" (McIntosh and Luengo-Fernandez, 2006).

"Economic evaluation is *a set of tools* that can be used to inform which interventions should be funded." (Revill, et al, 2015).

"Economic evaluation is *a set of formal analytical techniques* that provide systematic information about costs and benefits of alternative options, and can thereby assist in priority-setting." (Bergmo, 2009).

"Economic evaluation is *a general methodological approach* to assess the relative worth of various activities, in cases where such assessments are not performed by the normal market mechanisms." (Neymark, 2005).

"Economic evaluation is *a set of formal quantitative methods* to capture the outcomes and costs of alternate intervention strategies." (Qizilbash, 2008).

"Economic evaluation in healthcare is *a set of formal, quantitative methods* used to compare two or more healthcare interventions with respect to their resource use and expected outcomes." (Price and Christenson, 2003).

"An economic evaluation is *a method of gathering standardized, quantitative data* of the estimated costs of health benefits arising from the available treatment interventions." (Button, et al, 2009).

Some researchers define EE through detailing the steps of the evaluation process: identification, measurement, valuation and comparison of costs and outcomes.

"Economic evaluation methods provide a systematic way *to identify, measure, value, and compare* the costs and consequences of various programs, policies, or interventions." (Dunet, 2012).

"Full economic evaluation is a comparative analysis of alternative options/programmes that involves *identification, measurement, and valuation* of both costs and outcomes, and establishes the difference in costs in relation to difference in outcomes in an incremental fashion." (Gospodarevskaya and Westbrook, 2014).

"Economic evaluation is a technique by which one *identifies, measures, values and compares* different alternatives in terms of their costs and consequences." (Mateus and Coloma, 2013).

"Economic evaluation is a methodology that allows researchers *to identify, measure, rate and compare* different alternatives in terms of their costs and health consequences." (Rodríguez-Blázquez, et al, 2015).

"Economic evaluation - use applied analytical techniques *to identify, measure, value and compare* the costs and outcomes of alternative interventions. Types of economic evaluations include cost-benefit, cost-effectiveness, cost-efficiency evaluations." (UNAIDS, n.d.).

Some definitions introduce EE through a list of methods involved in the framework, e.g. cost-benefit analysis, cost-effectiveness analysis, cost-utility analysis, etc.

"Economic evaluation: this is the comparative assessment of interventions to improve health in terms of both their costs and their benefits. The different forms of economic evaluation (*cost-benefit analysis, cost-consequence analysis, cost-effectiveness analysis, cost-minimization analysis, and cost-utility analysis*) all share the same framework. Each evaluates cost in the same way but they differ from each other in the way that the outcomes or benefits of the interventions are included in the evaluation. This affects the types of questions that each technique can answer." (De Salazar, et al, 2007).

"Economic evaluation. An economic study design that allows the consequences of different interventions to be measured against a single outcome, usually in 'natural' units (for example, life-years gained, deaths avoided, heart attacks avoided, cases detected). Alternative interventions are then compared in terms of cost per unit of effectiveness." (NICE, 2009).

As it was mentioned, some authors follow Drummond's definition very closely or with minimal editorial changes.

"Health economic evaluation is a systematic, comparative analysis of competing decisions or alternative courses of action in terms of both their costs (use of resources) and their consequences (health benefits)." (O'Reilly, et al, 2011).

"Economic evaluation: comparison of two or more alternative courses of action in terms of both their costs and their consequences; economists usually distinguish several types of economic evaluation differing in how consequences are measured." (CDC, n.d.).

"An economic evaluation is a comparative study that examines the difference between costs and benefits between two or more options and is used in the estimation of cost effectiveness." (Sutton, et al, 2015).

"Economic evaluation (economic appraisal). The comparison of alternative courses of action in terms of their costs and consequences, with a view to making a choice." (BMJ, n.d.)

> "Economic evaluation—relating the costs and benefits of alternative ways of delivering health care." (Kernick, 2003).

Some definitions, although referring to health economics or economics as a whole, elucidate certain aspects of economic evaluation:

> "It analyses the costs and benefits of improving patterns of resource allocation." (Samuelson, 1976).

> "… one role of health economics is to provide a set of analytical techniques to assist decision making, usually in the health care sector, to promote efficiency and equity." (Shiell et al., 2002).

To conclude the review of the EE definitions, it should be noted that in almost all analyzed groups an attempt was made to add healthcare-specific flavour to definitions. That is achieved by using such terms/concepts as health economic evaluation, interventions, health effects, etc.

## 3   CONCLUDING REMARKS

EE is a dynamically advancing knowledge area and its definition requires an update. We found and analyzed 60 non-identical EE definitions, and grouped them into 8 clusters with common features.

Future research will be focused on the in-depth analysis of the revealed common features and developing a comprehensive EE definition.

**Note 1.** Data for the graphs shown in Figure 2 were obtained by searching the EBSCOhost Online Research Databases (https://www.ebscohost.com). The graph shows two series of bars pertaining to the results of two searches. One search was conducted in the Medline Complete database and another one was conducted concurrently in four health-related business-oriented databases: Business Source Complete, Health Business Elite, Health Policy Reference Center. Both searches used the following key words: "health care" AND ("economic evaluation" OR "cost benefit" OR "cost effectiveness" OR "cost consequence" OR "cost minimization" OR "cost minimisation" OR "cost utility"). Searches were limited to the titles of the papers (not full text). Searches were conducted in April 2016 in peer-reviewed academic journals. No limitation was set on the language of publications or geographic area. Each graph bar indicates a number of papers published during a five-year period with the whole timeline spanning six decades (from 1960 to 2020). The data for the period 2015-2020 have been projected based on the actual number of papers for the period from January 2015 to March 2016. Extrapolation used a conservative assumption that the monthly averages for the observed part of the period will be retained for the rest of the period.

Figure 2 shows that the number of EE articles indexed in the Medline database has the third order polynomial growth. Articles in the Medline database reflect studies of biomedical (disease-specific) type. The second search in the business-oriented databases represented studies in health services, health policy and evaluation methodology. Although the second type of studies also demonstrates growth, it is far less steep especially in the two most recent decades.

Although searched EBSCO databases index thousands of journals and thus provide a representative picture of the publication dynamics on the topic, there is no warranty or suggestion of the completeness of the data. Also, it should be noted that there is some overlap between the lists of journals in searched databases.